\documentclass[a4paper]{article}
\usepackage{geometry}
\usepackage[utf8]{inputenc}
\usepackage[T1]{fontenc}

\usepackage{graphicx,xcolor}
\usepackage{amsmath,amssymb,amsfonts}
\usepackage{url}

\usepackage{cite}
\usepackage{xcolor}

\usepackage{graphicx}
\usepackage{amsmath}

\usepackage{algorithm}
\usepackage{algpseudocode}
\algblockdefx{When}{EndWhen}[1]{\textbf{when} #1 \textbf{do}}{\textbf{end when}}

\usepackage[caption=false,font=footnotesize]{subfig}
\usepackage{url}

\usepackage{multirow}

\usepackage{comment}

\title{A State Transfer Method That Adapts to Network Bandwidth Variations in Geographic State Machine Replication\thanks{This work was supported by JSPS KAKENHI Grant Number JP18K18029.}}

\author{Tairi Chiba, Ren Ohmura, and Junya Nakamura\thanks{Corresponding author: junya[at]imc.tut.ac.jp}}

\date{Toyohashi University of Technology, Japan}

\begin{document}

\maketitle

\begin{abstract}

We present a new state transfer method for geographic State Machine Replication (SMR) that dynamically allocates the state to be transferred among replicas according to changes in communication bandwidths.
SMR is a method that improves fault tolerance by replicating a service to multiple replicas.
When a replica is newly added or is recovered from a failure, the other replicas transfer the current state of the service to it.
However, in geographic SMR, the communication bandwidths of replicas are different and constantly changing.
Therefore, existing state transfer methods cannot fully utilize the available bandwidth, and their state transfer time becomes long.
To overcome this problem, our method divides the state into multiple chunks and assigns them to replicas based on each replica's bandwidth so that the broader a replica's bandwidth is, the more chunks it transfers.
The number of assigned chunks is dynamically updated based on the currently estimated bandwidth.
The performance evaluation on Amazon EC2 shows that the proposed method reduces the state transfer time by up to 47\% compared with the existing one.

\end{abstract}

\section{Introduction}
\label{sec:introduction}

\emph{State Machine Replication (SMR)} \cite{Schneider1990} is a method that improves fault tolerance of a service.
SMR replicates a service to multiple servers called \emph{replicas}, which agree on the order of request processing among them to keep the state of each replica the same.
This method allows other replicas to continue the service even if some of them fail.
Furthermore, performing SMR with multiple geographically separated replicas allows the service to resist large-scale disasters such as earthquakes.
Such a method is called \emph{geographic SMR}.

In SMR, a replica that is newly added to replication or is recovered from faults, which is called a \emph{recovery replica}, retrieves the latest state of the service from the others called \emph{transfer replicas} to keep its state the same as others.
This process is called \emph{state transfer}, which is important because we cannot avoid faults of replicas in a long-run replication, and this process allows SMR to handle such faults without stopping the entire system.
Schneider introduced a basic state transfer method in which a recovery replica obtains the whole state from a single transfer replica \cite{Schneider1990}.
After that, Bessani et al. proposed the Collaborative State Transfer (CST) protocol \cite{Bessani2013}, which can mitigate the decrease in the request processing performance during a state transfer. 
In this method, each transfer replica sends an equally sized part of the current state to a recovery replica to reduce the state transfer time.
Existing state transfer methods assume that replicas are deployed in the same data center, in which the replicas are connected through high-speed and stable LAN.

\begin{figure*}[t]
\centering
    \vspace{-3mm}
    \subfloat[North Virginia (Group A)]{
        \includegraphics[width=70mm]{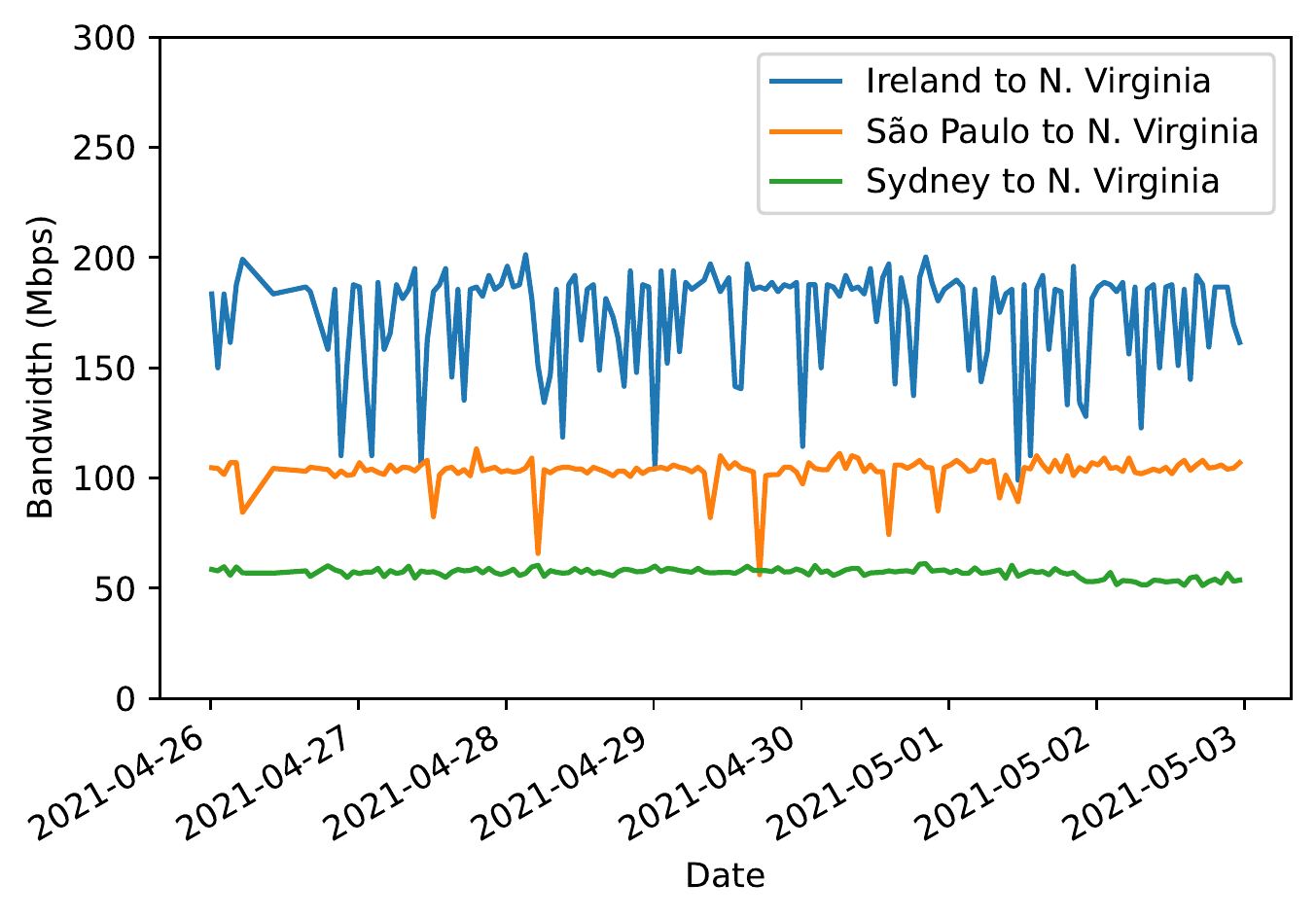}
        \label{fig:bandwidth-change-group-a-virginia}
    }
    \subfloat[London (Group B)]{
        \includegraphics[width=70mm]{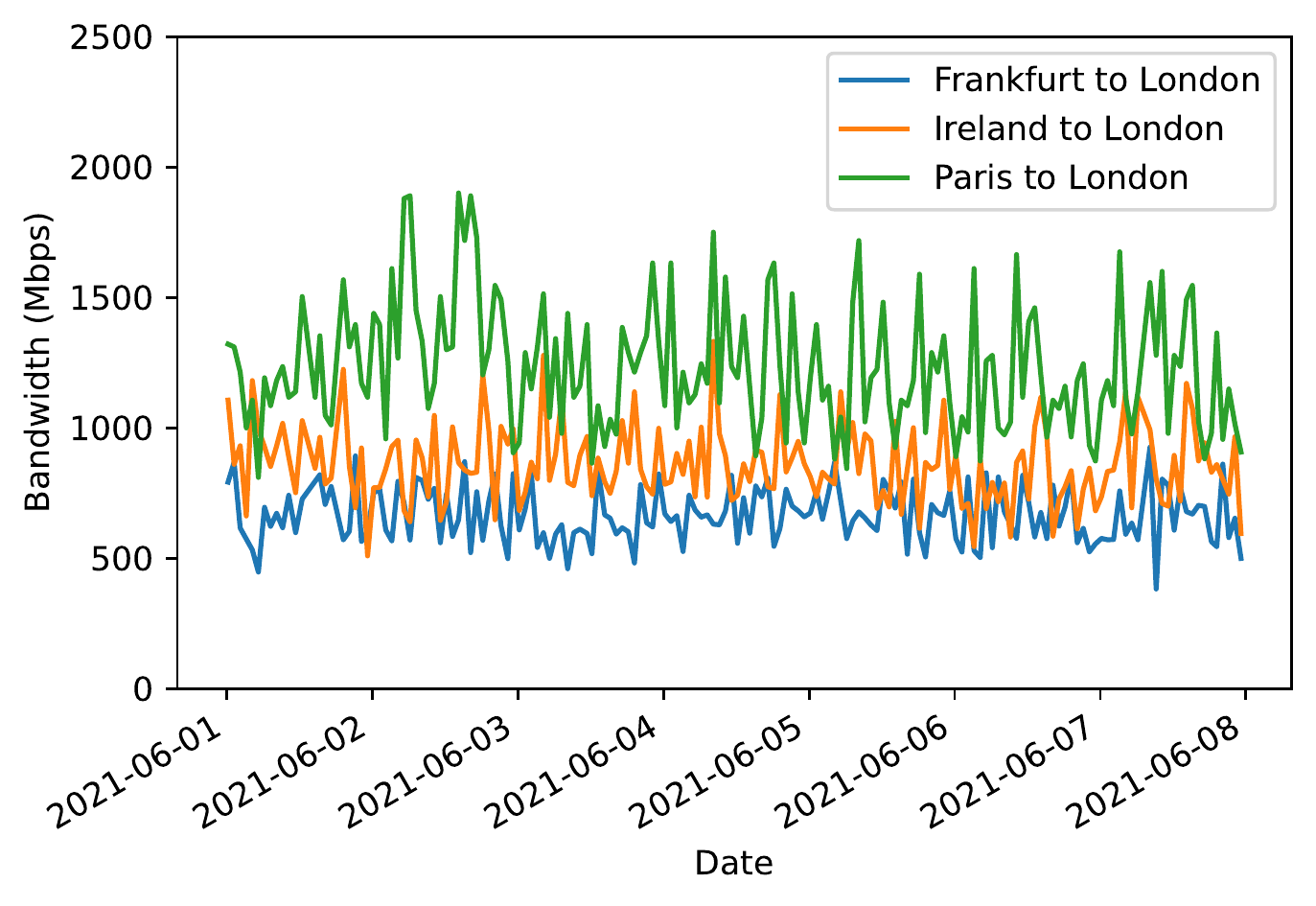}
        \label{fig:bandwidth-change-group-b-london}
    }
    \caption{Time variation of communication bandwidth in Amazon EC2 regions.}
    \label{fig:bandwidth-change}
\end{figure*}

On the other hand, in geographic SMR, replicas communicate with each other over WAN, so the characteristics of the communication environment are different from those of conventional SMR.
Figure \ref{fig:bandwidth-change} shows the change in the communication bandwidth over seven days between Amazon EC2 regions, commonly used in geographic SMR.
The communication bandwidth was measured hourly using iperf 3.1.3.
We deploy two groups of replicas, called Groups A and B, into regions.
Group A deploys replicas in four different continents: North Virginia, Ireland, São Paulo, and Sydney, while Group B deploys replicas in European regions: London, Frankfurt, Ireland, and Paris.
In Group A (Fig.~\ref{fig:bandwidth-change}\subref{fig:bandwidth-change-group-a-virginia}), the communication bandwidth differs greatly among replicas.
The degree of change in the communication bandwidth varies from region to region, with little changes in São Paulo and large changes in Ireland.
On the other hand, in Group B(Fig.~\ref{fig:bandwidth-change}\subref{fig:bandwidth-change-group-b-london}), the difference in the average communication bandwidth among replicas is small, but changes are very large for all regions.
Thus, in a geographic SMR environment where the available communication bandwidth differs for each replica and changes frequently and largely, the existing state transfer methods cannot transfer the state efficiently.

In this paper, we propose a state transfer method that adapts to network bandwidth variations in geographic SMR.
The proposed method has the following features:

\textbf{Chunk division and bandwidth-based allocation:}
The proposed method divides the replicated service state into several small \emph{chunks} and changes the number of chunks allocated to a replica so that a replica with a broader communication bandwidth transfers more chunks than others.
Ideally, the state should be strictly divided based on communication bandwidths, but for this purpose, a recovery replica needs to know the current state size in advance.
However, since geographic SMR generally has high communication delays \cite{ThousandEyes2019}, increasing the number of communications will increase the state transfer time and complicate the state transfer process.
Therefore, the proposed method divides the whole state into small $N$ chunks, where $N$ is the predefined value, and a recovery replica requests, as a partial state, $m$ chunks so that $m$ is close to the ratio of the communication bandwidth between each replica.

\begin{figure}[t]
\centering
\includegraphics[width=70mm]{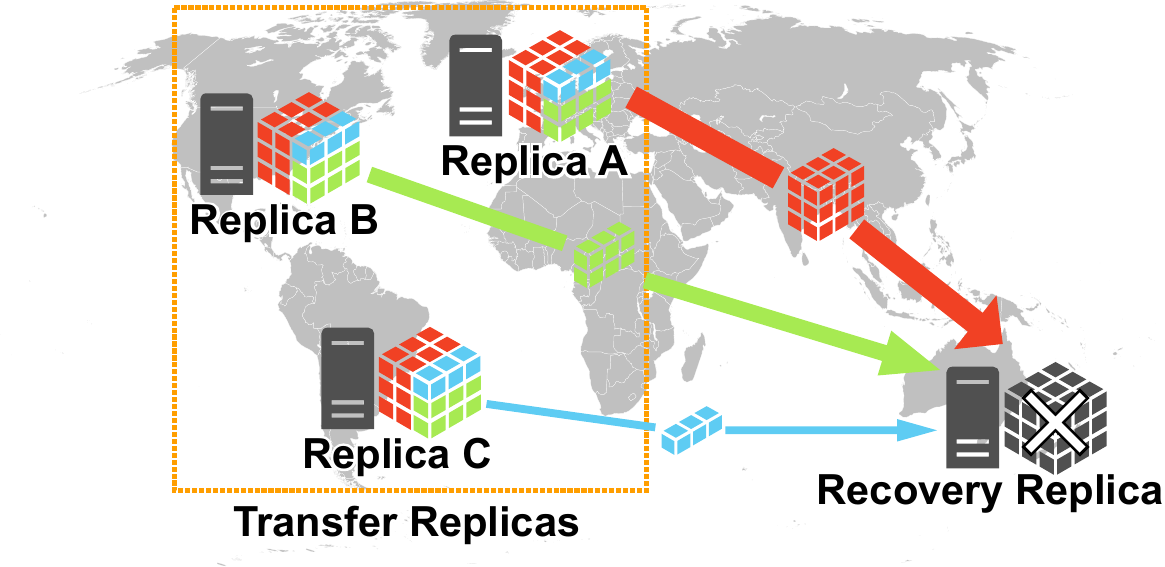}
\caption{An example of state transfer with the proposed method.}
\label{fig:proposed-fig}
\end{figure}

Figure \ref{fig:proposed-fig} shows an example of state transfer with the proposed method.
In Fig.~\ref{fig:proposed-fig}, three transfer replicas A, B, and C transfer chunks to the recovery replica.
Replica A and C have the broadest and narrowest bandwidth to the recovery replica, respectively.
The recovery replica estimates the communication bandwidth with each replica and requests the number of chunks close to the ratio of communication bandwidths, e.g., 18 chunks to replica A and 3 chunks to replica C.
The transfer replicas divide the state into chunks and transfer the chunks requested by the recovery replica.

\textbf{Updating allocated chunks according to changes in the communication bandwidth:}
In the proposed method, a recovery replica always measures the communication bandwidth with each replica based on the SMR message reception speed.
Note that this measurement is passive; thus, there is no overhead.
When the ratio of the communication bandwidth between replicas changes, the recovery replica periodically reallocates chunks based on the new ratio to adapt to dynamic changes in the communication bandwidth.

\textbf{Per-chunk hashing and efficient integrity verification:}
When a recovery replica receives a service state, it needs to verify that the state is really correct to avoid applying a false state from a malicious replica.
The proposed method efficiently verifies the integrity of the received state by assigning and verifying the integrity per chunk.

To show that the proposed method reduces the state transfer time of geographic SMR, we build a geographic SMR system on Amazon EC2 and evaluate the state transfer time of the proposed method and existing methods.
The experimental results show that the proposed method can reduce the state transfer time by up to 47\% compared to the CST protocol in the environment with a large difference (Group A) and large changes (Group B) in communication bandwidth.

\section{Related work}
\label{sec:related-work}

\subsection{State Machine Replication}

SMR \cite{Schneider1990} is a method of replicating a service to multiple servers (replicas) and processing requests in the same order to keep the replicas in the same state; as a result, a service can improve fault tolerance of the service.
Although SMR requires that all replicas receive requests in the same order, the communication delay between replicas is different in the actual network.
Even if a client sends a request to each replica at the same time, the request arrival time differs among replicas.
In such a case, a distributed protocol called Total Order Broadcast or Atomic Broadcast \cite{D'efago2004} can ensure that every message is delivered to all participants in a group in the same order.
We can realize SMR by using Total Order Broadcast to deliver requests to replicas.

We can classify SMR protocols into Crash Fault-Tolerant SMR (CFT-SMR) and Byzantine Fault-Tolerant SMR (BFT-SMR) according to their target failure model.
CFT-SMR resists crash faults in which a faulty replica does not operate at all after the failure, while the BFT-SMR resists Byzantine fault \cite{Lamport1982} in which a faulty replica behaves arbitrarily without following the protocol after the failure.
There are several protocols for CFT-SMR, such as Paxos \cite{Lamport1998}, Raft \cite{Ongaro2014}, and ZAB \cite{Junqueira2011}, and for BFT-SMR, such as PBFT \cite{Castro1999}, Zyzzyva \cite{Kotla2010}, and BFT-SMaRt \cite{Sousa2012}.
Geographic SMR places replicas at a large distance from each other and tolerates large-scale disasters such as earthquakes.
Since geographic SMR has different characteristics from conventional SMR, there are several SMR protocols for geographic SMR to improve the performance by focusing on the latency between replicas \cite{Mao2008,Zhao2018} and the non-uniformity of the performance of each replica \cite{Zhao2018}, as well as optimizing the replica placement based on the latency \cite{Numakura2019b}.
The proposed state transfer method is designed for geographic SMR and can be applied to both CFT-SMR and BFT-SMR. 

\subsection{State Transfer of SMR}

In SMR, the latest state is transferred to the failed replica to restore it to replication as a normal replica \cite{Schneider1990}.
Since the request processing performance degrades during state transfer, Bessani et al. proposed a fast state transfer method called CST protocol \cite{Bessani2013}, in which multiple replicas cooperate to transfer divided states to reduce the transfer time.

CST protocol operates as follows.
The CST protocol expresses the state of each replica to be transferred by the \emph{checkpoint} and the \emph{log}.
The checkpoint is a copy of the current service state obtained periodically, and the log is a list of requests executed after the latest checkpoint was obtained.
In the CST protocol, the replica that wants to get the latest state (\emph{recovery replica}) requests state transfer from all other replicas (\emph{transfer replica}), which have the latest state.
When the state is requested, a transfer replica transfers the entire checkpoint, and the others transfer different parts of the equally divided log at the same time.
In the case of BFT-SMR, to verify the integrity of the received information, the transfer replicas also send the hash of the checkpoint and the divided log.
Whether the transfer replica transfers the checkpoint or the log and which part of the split log is transferred are statically determined from the replica IDs assigned at the start of each transfer replica.
If the recovery replica receives the checkpoint and all parts of the log and, in the case of BFT-SMR, the verification by hash also succeeds, then the recovery replica applies the state to itself and completes the state transfer.

\section{Proposed Method}
\label{sec:proposed-method}

Here, we describe the proposed state transfer method.
We give the overview of this method in Sect.~\ref{sec:overview} and show the details of the method in Sect.~\ref{sec:details}.
Finally, we prove the correctness of this method in Sect.~\ref{sec:correctness}.

\subsection{Overview}
\label{sec:overview}

\begin{figure}[t]
    \centering
    \includegraphics[width=70mm]{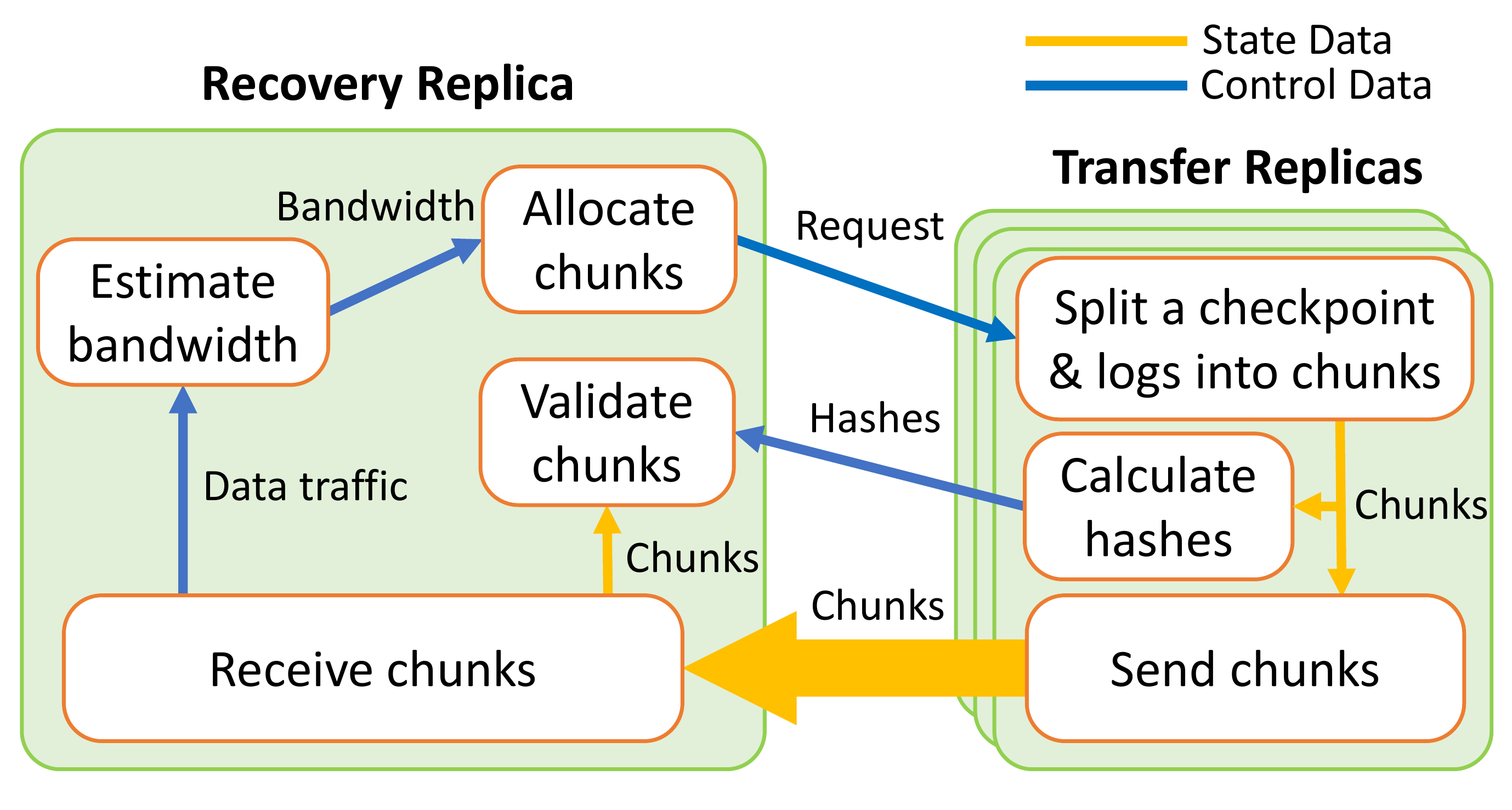} 
    \caption{Overview of the proposed method}
    \label{fig:proposed-block}
\end{figure}

Figure \ref{fig:proposed-block} shows the overview of the proposed method.
The proposed method internally represents the service state by the checkpoint and the log as in the CST protocol, but during state transfer, the state is represented as the binary that concatenations them.
Furthermore, the state is divided into $N$ chunks of the same size and transferred in these units.
In this way, the amounts of data transferred by each replica can be easily adjusted.

A recovery replica assigns $N$ chunks to the transfer replicas at the start of state transfer and requests the transfer of chunks to them.
When a transfer replica receives the chunk transfer request, it divides the checkpoint and the log into chunks.
The transfer replica then sends the chunks and the hashes of them, which will be used to verify the integrity of the chunks, to the recovery replica.

When the recovery replica receives a chunk from the transfer replica, it calculates a hash of the received chunk and compares the hash it calculated with the hash it received from the transfer replica to exclude the fake chunk sent by a Byzantine replica.
This verification proceeds in parallel with the sending and receiving process for state transfer.
This reduces the state transfer time and enables early detection of fake chunks and their re-request.

The recovery replica always records the reception rate of chunks from each transfer replica during state transfer and estimates the communication bandwidth based on the rate.
Based on the estimated bandwidth, the recovery replica adjusts the chunk allocation of the transfer replicas so that the replica with the broader communication bandwidth than others transfers more chunks.
In this way, the proposed method adapts to dynamic changes in the communication bandwidth.

When the recovery replica receives all chunks and completes the verification, it restores the checkpoint and the log from the chunks and applies them to the replica along with the received log during state transfer to restore the latest service state and ends state transfer.

\subsection{Details}
\label{sec:details}

Here, we describe the details of the proposed method.
First, we introduce some notations to describe the method formally.
In the proposed method, all replicas except the recovery replica $r$ behave as transfer replicas.
We denote the set of all transfer replicas by $T$.
The recovery replica repeatedly requests a set of chunks from the transfer replica every $I$ seconds.
The $i$-th chunk set that the recovery replica $r$ requests from the transfer replica $t$ is denoted by $C_t^i$, and the set of all chunks that represents the latest service state is denoted by $C_{all}$.
The average communication bandwidth with the transfer replica $t$ estimated by the recovery replica between the ($i-1$)-th and $i$th times is denoted as $w_t^i$, and the sum of the communication bandwidth with all transfer replicas is denoted as $w_{all}^i=\sum_{j \in T}w_j^i$.
In the case of $i=0$, a recovery replica cannot estimate the bandwidth; thus, we assume $w_t^0=1$ for every transfer replica $t$.

In BFT-SMR, Byzantine replicas may transfer fake chunks to a recovery replica as transfer replicas.
To avoid the effects of such attacks, the recovery replica collects hashes of each chunk and verifies the integrity of the received chunks.\footnote{In CFT-SMR, a faulty replica just stops. Therefore, a recovery replica does not need to take care of such attacks.}
The set of hashes of all chunks is denoted as $H_{all}$, the set of hashes that a recovery replica verifies the integrity of their chunks is denoted as $H$, and the hash function is denoted as $\mathit{Hash}(x)$.

Algorithm \ref{alg:recovery-replica} shows the pseudocode of a recovery replica for BFT-SMR.
When state transfer is required, a recovery replica $r$ starts two tasks: $T_1$ to request chunks and $T_2$ to receive chunks.
First, task $T_1$ requests hashes $H_{all}$ of all chunks for all the transfer replicas $T$ (lines \ref{algl:t1_hash_req_begin}--\ref{algl:t1_hash_req_end}).
Next, $r$ requests a set of chunks $C_{all}$ from $T$ every $I$ seconds until all chunks $C_t^i$ are received (lines \ref{algl:t1_chunk_req_begin}--\ref{algl:t1_chunk_req_end}).
Here, $r$ determines $C_t^i$ so that it satisfies the conditions
\begin{equation}
    \label{eq:chunk_cond_1}
    |C_t^i| = (N - |C|) \cdot w_t^i/w_{all}^i
\end{equation}
\begin{equation}
    \label{eq:chunk_cond_2}
    C \cup \bigcup_{j \in T} C_j^i = C_{all}, 
\end{equation}
where $C$ is the set of chunks that $r$ has received and verified.
In the case $|C_t^i|=0$ because of narrow bandwidth, a recovery replica allocates one chunk, which overlaps with $C_a^i$ of another transfer replica $a$.\footnote{This is because some data transfer is necessary to passively estimate the communication bandwidth.}

Task $T_2$ continues to receive chunks from transfer replicas until it receives all chunks $C_{all}$ (lines \ref{algl:t2_chunk_req_begin}--\ref{algl:t2_chunk_req_end}).
After receiving a chunk $c$ that has not been verified, a recovery replica $r$ verifies the integrity of the chunk (lines \ref{algl:t2_chunk_verify_begin}--\ref{algl:t2_chunk_verify_end}).
The verification succeeds if $r$ receives the same hash from $f+1$ transfer replicas and $h$ matches hash $h' = \mathit{Hash}(c)$ that is locally computed in $r$ from $c$.
If the verification succeeds, $r$ includes $c$ in $C$.
When the verification fails, recovery replica $r$ requests $c$ again to another transfer replica different from the previous sender of $c$ in the next chunk request.
After receiving all the chunks, $T_2$ combines $C$ to get the checkpoint and the log and uses them to restore the latest state (lines \ref{algl:t2_state_recovery_begin}--\ref{algl:t2_state_recovery_end}).
When the recovered replica is up-to-date, the proposed method terminates.

Algorithm \ref{alg:transfer-replica} shows the pseudocode of a transfer replica for BFT-SMR.
When a transfer replica $t$ receives a request for hashes of all chunks from a recover replica $r$, $t$ calculates $H$ from the latest checkpoint and the log and sends it to $r$ (lines \ref{algl:hash_transfer_begin}--\ref{algl:hash_transfer_end}).
On the other hand, when a received request wants a chunk set $C_t^i$, $t$ divides the checkpoint and the log into chunks and transfers the requested chunks to $r$ one by one (lines \ref{algl:chunk_transfer_begin}--\ref{algl:chunk_transfer_end}).

In the case of CFT-SMR, the following processes are changed.
First, in $T_1$, the request of $H$ (lines \ref{algl:t1_hash_req_begin}--\ref{algl:t1_hash_req_end} of Alg.~\ref{alg:recovery-replica}) is no longer needed.
Next, $T_2$ skips the integrity verification of the received chunk $c$ (lines \ref{algl:t2_chunk_verify_begin}--\ref{algl:t2_chunk_verify_end} of Alg.~\ref{alg:recovery-replica}) and simply adds $c$ to $C$.
Finally, a transfer replica is never requested the transfer of $H$ (lines \ref{algl:hash_transfer_begin}--\ref{algl:hash_transfer_end} of Alg.~\ref{alg:transfer-replica}).

\begin{algorithm}[t]

\caption{The pseudocode of the recovery replica}
\label{alg:recovery-replica}

\begin{algorithmic}[1]
\small
\Statex \hspace{-2em} Initialization:
\State $C \gets \emptyset$
\State \textbf{Activate task} $T_1$, $T_2$
\vspace{2mm}

\Statex \hspace{-2em} Task $T_1$:
\State $i \gets 0$
\For{$t$ \textbf{in} $T$} \label{algl:t1_hash_req_begin}
    \State Request the set $H_{all}$ of all hashes to $t$
\EndFor \label{algl:t1_hash_req_end}
\While{$C \neq C_{all}$} \label{algl:t1_chunk_req_begin}
    \For{$t$ \textbf{in} $T$}
    \State Calculate $C_{t}^i$ from Equations.~\eqref{eq:chunk_cond_1} and \eqref{eq:chunk_cond_2}
        \State Request $C_{t}^i$ to $t$
    \EndFor
    \State Wait $I$ seconds
    \State $i \gets i + 1$
\EndWhile \label{algl:t1_chunk_req_end}
\vspace{2mm}

\Statex \hspace{-2em} Task $T_2$:
\When{a chunk set $C_t \subset C_{all}$ is delivered from a recovery replica $t \in T$}
    \For{$c$ \textbf{in} $C_t \setminus C_{all}$} \label{algl:t2_chunk_req_begin}
        \If{it has already received the same $f+1$ hash $h$ of chunk $c$ and $h$ matches $h' = \mathit{Hash}(c)$}
        \label{algl:t2_chunk_verify_begin}
            \State $C \gets C \cup \{c\}$
        \EndIf \label{algl:t2_chunk_verify_end}
    \EndFor \label{algl:t2_chunk_req_end}
    \If{$C = C_{all}$}
        \State Combine $C$ and get checkpoint $a$ and log $L$ \label{algl:t2_state_recovery_begin}
        \State Apply checkpoint $a$
        \State Process log $L$ and the log received during state transfer in order of oldest to newest \label{algl:t2_state_recovery_end}
        \State \textbf{Terminate}
    \EndIf
\EndWhen

\end{algorithmic}
\end{algorithm}

\begin{algorithm}[t]

\caption{The pseudocode of the transfer replica}
\label{alg:transfer-replica}
\begin{algorithmic}[1]
\small

\When{the hash set $H_{all}$ of all chunks is requested from a recovery replica $r$} \label{algl:hash_transfer_begin}
    \State Transfer $H_{all}$ to $r$
\EndWhen \label{algl:hash_transfer_end}
\vspace{2mm}

\When{a chunk set $C_t^i$ is requested from a recovery replica $r$} \label{algl:chunk_transfer_begin}
    \State $C_t \gets C_t^i$
    \For{$c$ \textbf{in} $C_t$}
        \State Transfer $c$ to $r$
    \EndFor
\EndWhen \label{algl:chunk_transfer_end}

\end{algorithmic}
\end{algorithm}

\subsection{Correctness}
\label{sec:correctness}

We prove the correctness of this method in terms of the following \emph{safety} and \emph{liveness}.
\begin{itemize}
    \item \textbf{Safety:} The state applied to a recovery replica is the same as that of a non-faulty replica.
    \item \textbf{Liveness:} State transfer eventually terminates.
\end{itemize}

For safety, a recovery replica verifies the integrity of a chunk $c$ using identical hashes $h$ received from $f+1$ transfer replicas.
In addition, the recovery replica locally computes $h' = \mathit{Hash}(c)$ and compares it to $h$ to deal with a Byzantine transfer replica sending a fake chunk and the correct hash.
This ensures that the chunk is identical to that of at least one non-faulty transfer replica. 
Therefore, the state applied to the recovery replica is also the same as that of a non-faulty replica.

Next, we prove that liveness.
A recovery replica $r$ needs to receive all hashes and chunks and verify their integrity before terminating.
For receiving hashes, even if $f$ transfer replicas are faulty and send invalid hashes, there are still at least $f+1$ non-faulty transfer replicas; then, $r$ can receive $f+1$ identical hashes.
For receiving chunks, a Byzantine transfer replica does not send any chunks.
However, a recovery replica $r$ continues to request at least one chunk from each transfer replica.
Thus, even if the Byzantine transfer replica does not send any chunks, $r$ eventually receives all the chunks from non-faulty transfer replicas.
Therefore, state transfer eventually terminates.

\section{Performance Evaluation}
\label{sec:evaluation}

We build a geographic BFT-SMR system on Amazon EC2 using the open-source SMR library BFT-SMaRt 1.2\footnote{\url{https://github.com/bft-smart/library/releases/tag/v1.2}} and evaluate the state transfer time of the proposed method.

\subsection{Experimental Method}
We implement the proposed method and the CST protocol \cite{Bessani2013} in BFT-SMaRt as a state transfer method.
While \cite{Bessani2013} introduced various optimization techniques for the CST protocol, we implement only ``Optimizing CST'' that improves the state transfer time. 
In the CST protocol, the sizes of the checkpoint and logs sent by transfer replicas depend on the timing of the state transfer, but we assume that each transfer replica sends the partial state of the same size because this is the fastest situation for the CST protocol.

In addition, as a baseline, we implement a method that removes the dynamic bandwidth measurement function from the proposed method.
This method allocates chunks using the average communication bandwidth measured in advance.
This communication bandwidth between replicas is measured hourly for seven days using iperf 3.1.3.
Table \ref{tb:bandwidth} shows the measured communication bandwidth between each replica in two groups. 
In each table, the first column represents sending replicas and the first row represents receiving replicas.
In the following, we refer to this method as ``Premeasured BW.''

\begin{table}[t]
    \centering
    \caption{Premeasured inter-region communication bandwidth (Mbps).}
    \label{tb:bandwidth}
    \vspace{-3mm}
    \subfloat[Group A]{
        \begin{tabular}{|l|rrrr|}
            \hline
            & Sydney & São Paulo & N. Virginia & Ireland \\ \hline
            Sydney &    & 33.7  & 57.0  & 42.9 \\
            São Paulo & 33.3   &   & 102.2  & 64.5 \\
            N. Virginia & 56.6   & 103.0  &   & 174.3 \\
            Ireland & 42.9   & 64.4  & 173.3  & \\
            \hline
        \end{tabular}
        \label{tb:bandwidth-a}
    }
    
    \subfloat[Group B]{
        \begin{tabular}{|l|rrrr|}
            \hline
            & Ireland & London & Paris & Frankfurt \\ \hline
            Ireland &    & 857.8  & 578.1  & 420.8 \\
            London & 866.8   &   & 1219.6  & 667.2 \\
            Paris & 594.4   & 1234.4  &   & 1115.5 \\
            Frankfurt & 420.0   & 662.4  & 1113.6  & \\
            \hline
        \end{tabular}
    
        \label{tb:bandwidth-b}
    } 

\end{table}

In the experiment, we use four replicas: one recovery replica and three transfer replicas.
The geographic locations of replicas are Group A and Group B described in Section \ref{sec:introduction}.
Each replica uses a t3.xlarge instance.\footnote{\url{https://aws.amazon.com/ec2/instance-types/t3/}}
Table \ref{tb:aws-spec} shows the performance of the instance type.
In each replica, we use Docker 18.09.2 on Amazon Linux 2 and run BFT-SMaRt in the openjdk:14-alpine container.

Unless otherwise noted, the size of the state to transfer is 1000 MiB, the total number $N$ of chunks in the proposed method is 256, and the update interval $I$ of the chunk allocation in the proposed method is 1000 ms.
For the state transfer time, we use the average value of five measurements.

\begin{table}[t]
    \centering
    \caption{Specifications of t3.xlarge instance.}
    \label{tb:aws-spec} %
    \begin{tabular}{|c|r|} \hline
        Item & Specification \\ \hline
        CPU & Intel Xeon Platinum 8000 series \\
        \#vCPUs & 4 \\
        Memory (GiB) & 16.0 \\
        \begin{tabular}{c}
        Network burst \\
        bandwidth (Gbps)
        \end{tabular}
        & 5 \\
        \hline
    \end{tabular}
\end{table}

In the experiment, the size of the checkpoints is fixed, and each method divides only the checkpoint.
We refer to the size of the checkpoint as the \emph{state size}.
The log size was less than 1/1000 of the fixed state size in the preliminary experiment, and thus it does not affect the result.
We use SHA-512 as a hash function $\mathrm{Hash}(x)$ to verify the integrity of the chunks or the checkpoint and logs.

\subsection{Comparison of State Transfer Time}
\label{sec:compare-transfer-time}

The state transfer time of the proposed method is compared with the method using the CST protocol and premeasured BW.
We build geographic SMR systems for each group (A and B) and measure the state transfer time 12 times every 2 hours when each replica of the group becomes the recovery replica and obtain the average value.

\begin{figure*}[t]
    \centering
    \vspace{-3mm}
    \subfloat[Group A]{
        \includegraphics[width=70mm]{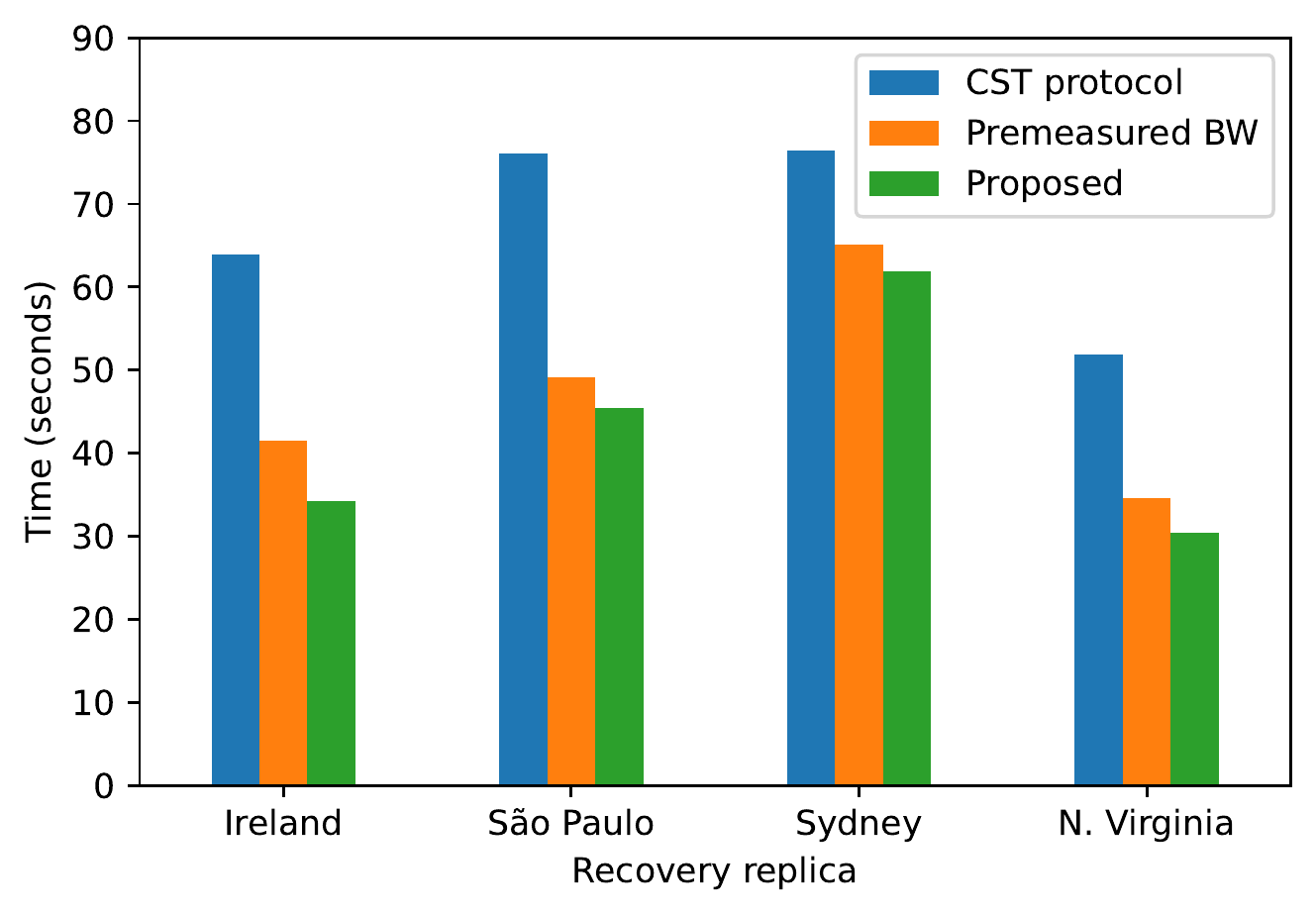}
        \label{fig:recovery-replica-and-transfer-time-group-a}
    }
    \subfloat[Group B]{
        \includegraphics[width=70mm]{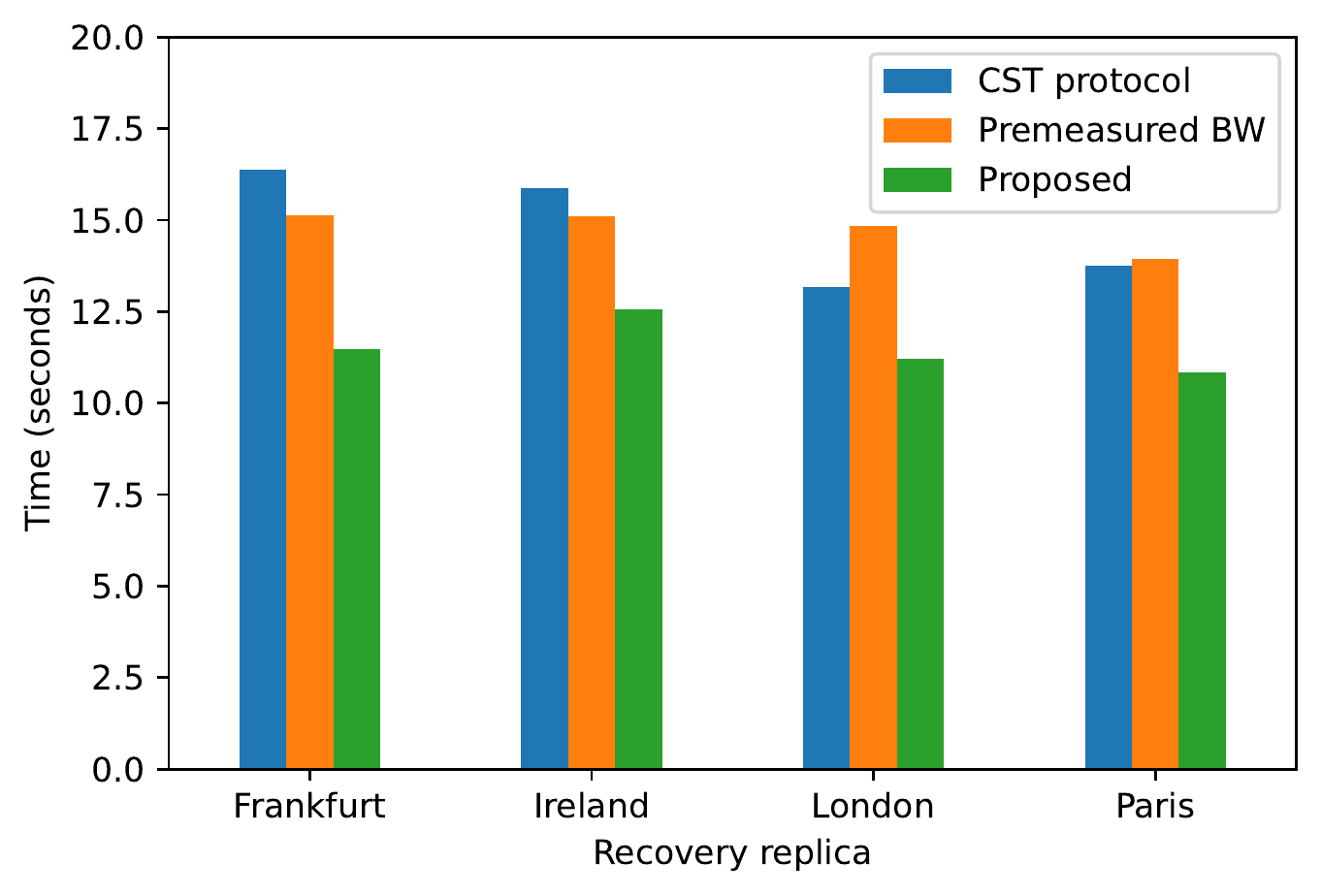}
        \label{fig:recovery-replica-and-transfer-time-group-b}
    }
    \caption{Location of the recovery replica and state transfer time when the state size is 1000 MiB.}
    \label{fig:recovery-replica-and-transfer-time}
\end{figure*}

First, we compare the average state transfer time for each recovery replica when the state size is 1000 MiB.
Figure \ref{fig:recovery-replica-and-transfer-time} shows the results.

In Group A, the state transfer time of the proposed method was the shortest for all the recovery replicas.
When the recovery replica was in Ireland, the difference between the proposed method and CST protocol was the largest and reduced by 47\%.
In contrast, the smallest reduction to the CST protocol was, when the recovery replica was in Sydney, at 19\%.
This difference is due to the difference in the communication bandwidth of the transfer replicas.
The average reducing rate of the proposed method for the CST protocol was 37\%.
On the other hand, the average reducing rate of the proposed method to the premeasured BW method was 10\%.
This is because the time variation of the communication bandwidth between replicas is small.

Unlike Group A, the state transfer time of the premeasured BW method was longer in Group B.
This result is because, as shown in Fig.~\ref{fig:bandwidth-change}\subref{fig:bandwidth-change-group-b-london}, the time variation of the communication bandwidth is large in Group B.
The results of Group B suggest the importance of dynamically responding to the time variation of the communication bandwidth.

\begin{figure*}[t]
    \centering
    \vspace{-3mm}
    \subfloat[North Virginia (Group A)]{
        \includegraphics[width=70mm]{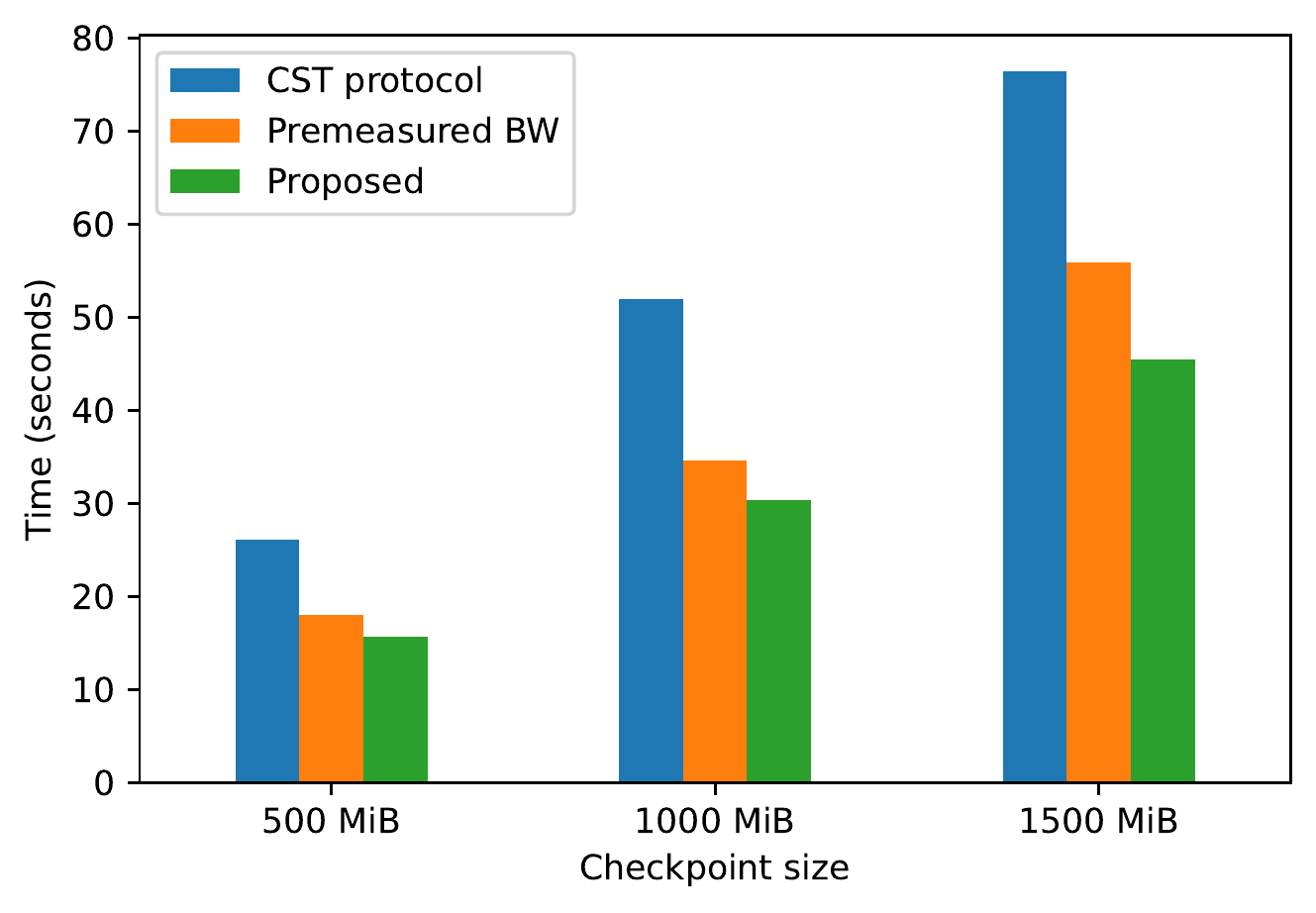}
        \label{fig:state-size-and-transfer-time-group-a}
    }
    \subfloat[Ireland (Group B)]{
        \includegraphics[width=70mm]{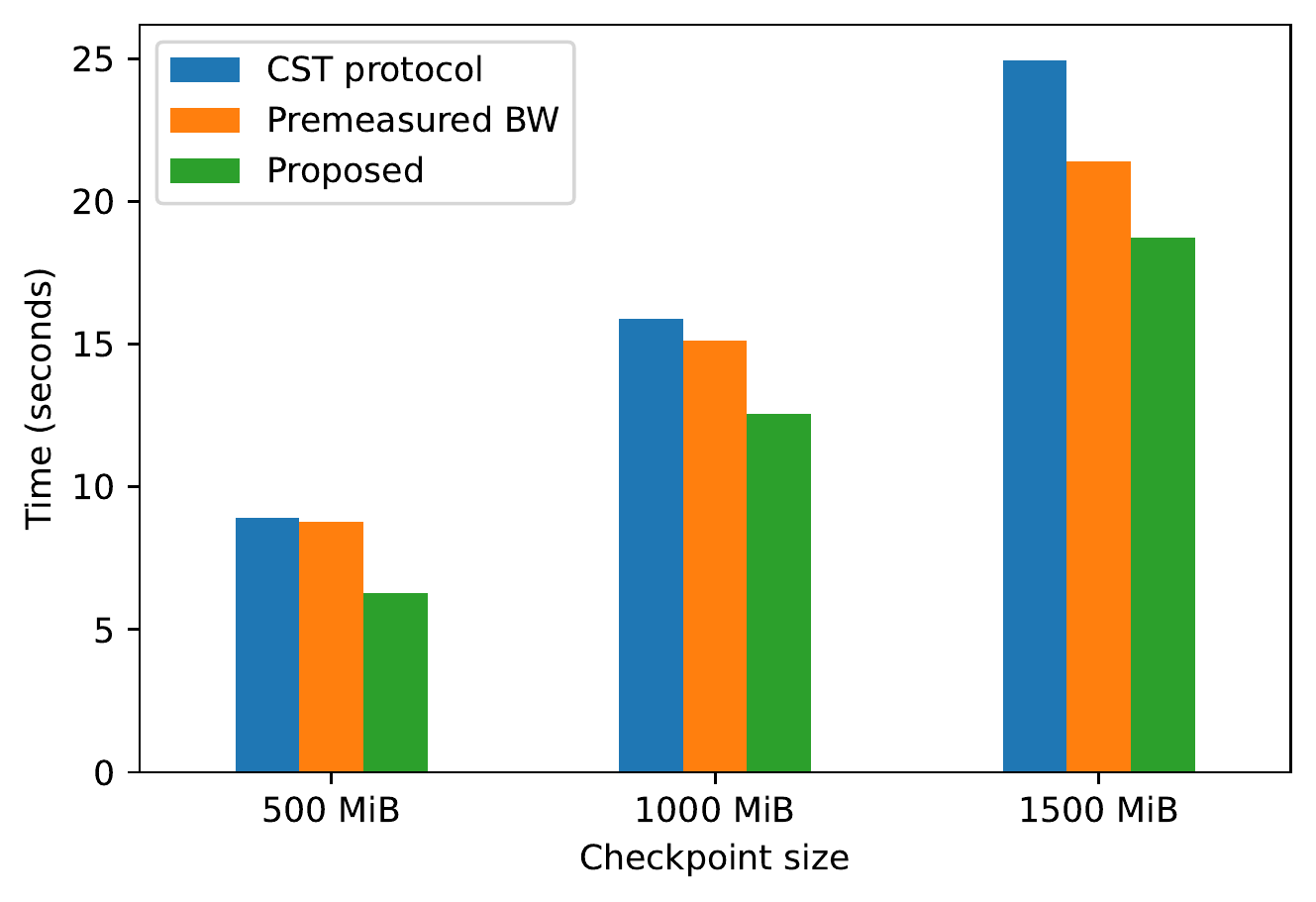}
        \label{fig:state-size-and-transfer-time-group-b}
    }
    \caption{State size and state transfer time.}
    \label{fig:state-size-and-transfer-time}
\end{figure*}

Next, we compare the effect of different state sizes on the state transfer time.
Figure \ref{fig:state-size-and-transfer-time} shows the average state transfer times for different state sizes of 500 MiB, 1000 MiB, and 1500 MiB in North Virginia in Group A and Ireland in Group B. 
In Group A, the effect of the state size was small: 40\% for the 500 MiB state, 42\% for the 1000 MiB state, and 41\% for the 1500 MiB state, compared to the CST protocol.
On the other hand, in Group B, the effect of state size was larger than in Group A: 30\% for the 500 MiB state, 21\% for the 1000 MiB state, and 25\% for the 1500 MiB state.
This result is because Group B was more strongly affected by the increase in transfer time due to the larger state size since the time variation of the communication bandwidth was larger than in Group A.

\subsection{State Transfer Time of Each Transfer Replica}
\label{sec:transfer-time-for-each-transfer-replicas}

\begin{figure*}[t]
    \centering
    \vspace{-3mm}
    \subfloat[North Virginia (Group A)]{
        \includegraphics[width=70mm]{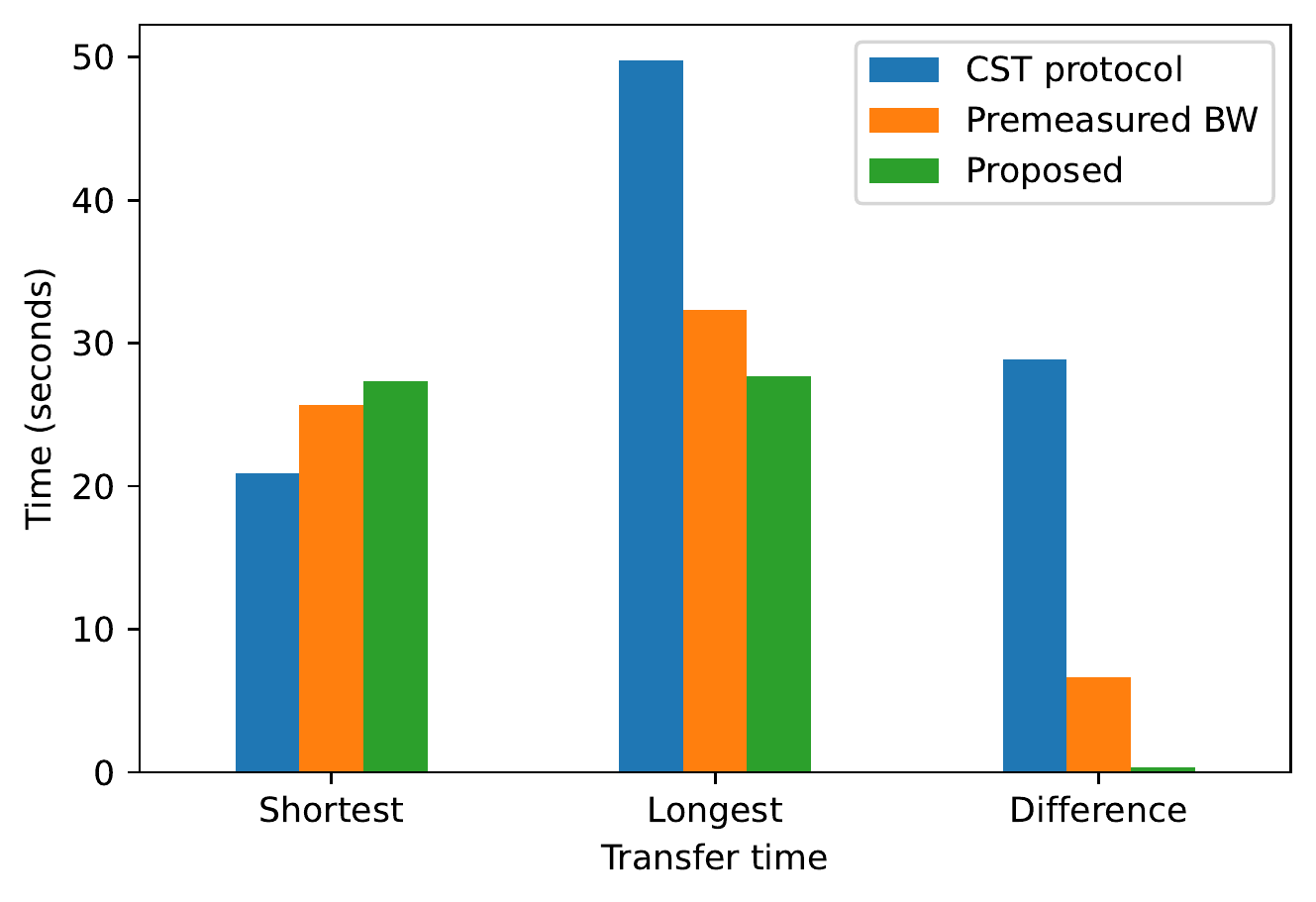}
        \label{fig:time-per-transfer-steps-a-virginia}
    }
    \subfloat[Ireland (Group B)]{
        \includegraphics[width=70mm]{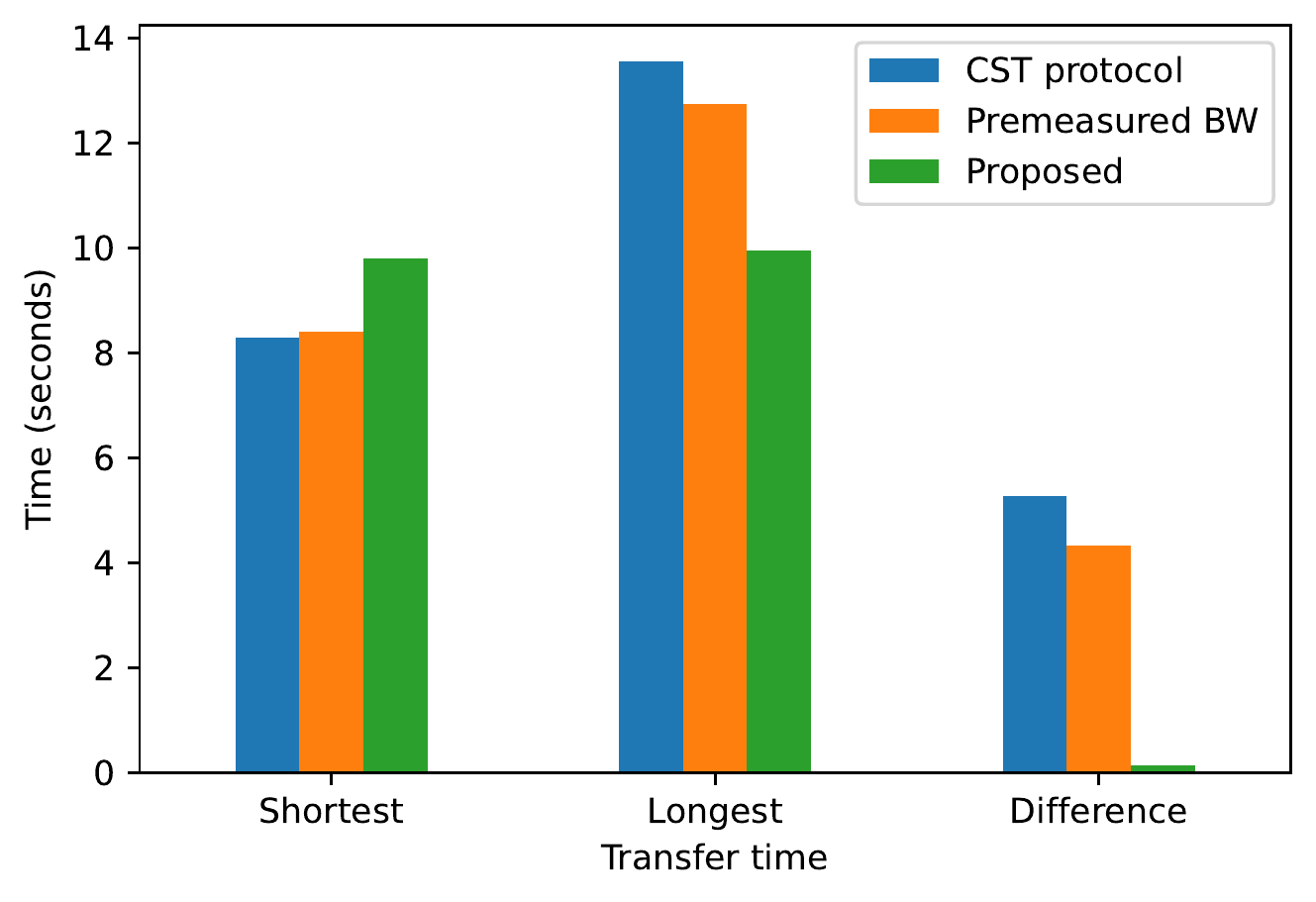}
        \label{fig:time-per-transfer-steps-b-ireland}
    }
    \caption{State transfer time for each replica. The horizontal axis shows the shortest and longest transfer times of the transfer replicas and their difference.}
    \label{fig:time-per-transfer-steps}
\end{figure*}

The proposed method reduces the total state transfer time by balancing the state transfer time of the transfer replicas.
We demonstrate the effect of balancing the state transfer time by comparing the state or chunk transfer times of the transfer replicas.
This experiment uses North Virginia in Group A and Ireland in Group B as the recovery replica.
Figure \ref{fig:time-per-transfer-steps} shows the state transfer time for each transfer replica.

As shown in Fig.~\ref{fig:time-per-transfer-steps}\subref{fig:time-per-transfer-steps-a-virginia} and Fig.~\ref{fig:time-per-transfer-steps}\subref{fig:time-per-transfer-steps-b-ireland}, the CST protocol and premeasured BW method have large difference in the state transfer time of the transfer replicas.
The recovery replica in North Virginia and Ireland has a difference of 2.38 times and 1.63 times with the CST protocol and 1.26 times and  1.54 times with the premeasured BW, respectively.
The CST protocol has no mechanism to adjust the difference in the communication bandwidth between replicas.
Therefore, the state transfer time tends to differ greatly for each replica.
This tendency is particularly remarkable in Group A.
The pre-measured BW method cannot follow the changes in the communication bandwidth while transferring chunks.
Therefore, the state transfer time tends to vary for each transfer replica in Group B, where the communication bandwidth changes frequently.
On the other hand, the difference of the proposed method is significantly small, 1.01 times in both cases.
The proposed method has features of dynamically adapting to both problems, and it works well.

\subsection{Effect of Different Hash Calculation Methods on State Transfer Time}
\label{sec:relation-between-hash-calc-way-and-transfer-time}

\begin{figure}[t]
    \centering
    \includegraphics[width=70mm]{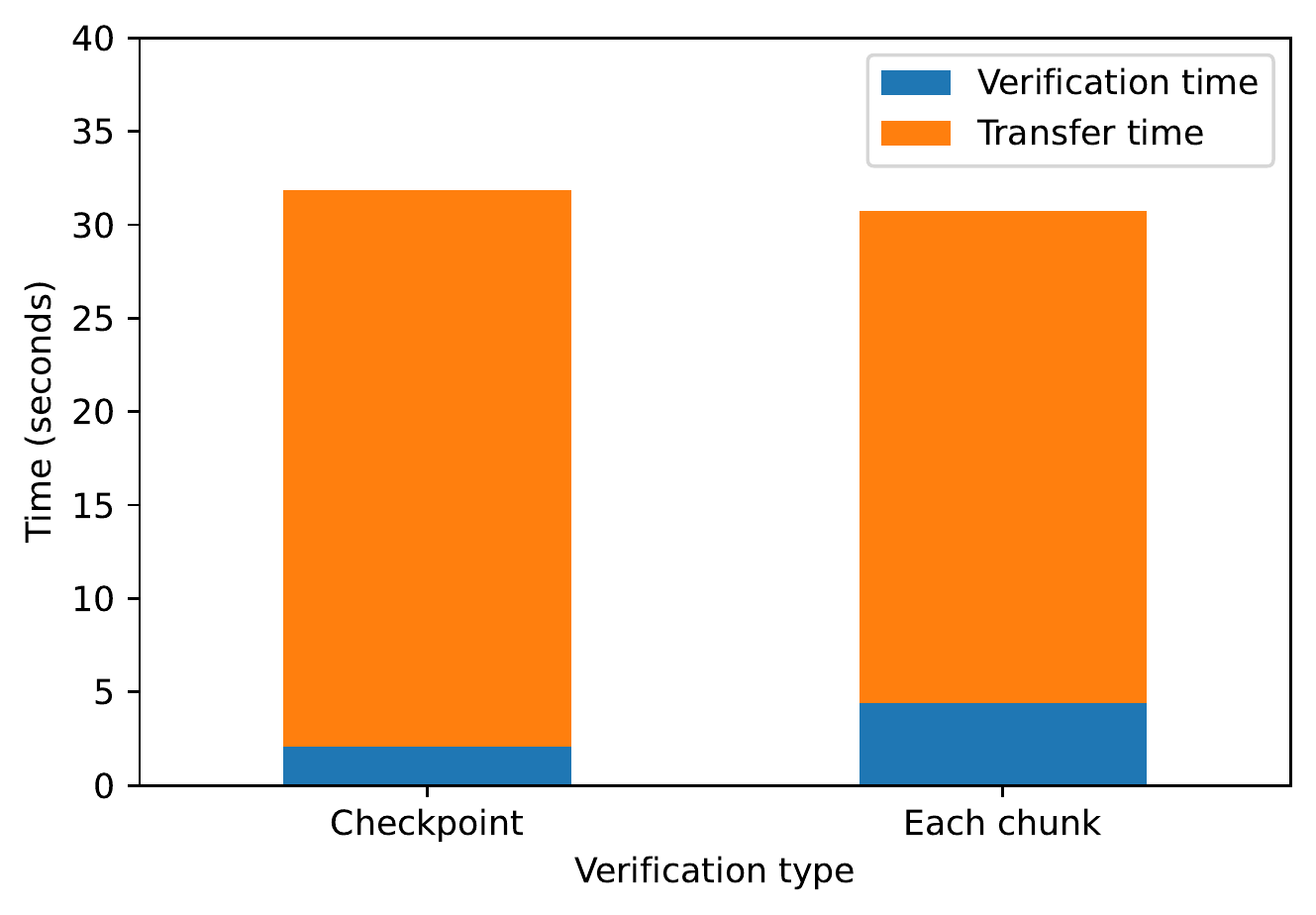}
    \caption{State transfer time for each hash verification method (North Virginia, Group A). The state transfer time includes the verification time.}
    \label{fig:relation-between-hash-calc-way-and-transfer-time-group-a-virginia}
\end{figure}

In the proposed method, a recovery replica calculates the hash of each chunk and verifies the hash in parallel with the chunk reception to reduce the transfer time.
To show the effect of this technique, we measured the difference in transfer time between the proposed method and the method that calculates the hash of the whole state.
In the latter method, all transfer replicas send the whole hash at the start of the transfer, and the recovery replica verifies the hash after the chunks are combined.
Figure \ref{fig:relation-between-hash-calc-way-and-transfer-time-group-a-virginia} shows the results measured by a recovery replica in North Virginia in Group A.

Figure \ref{fig:relation-between-hash-calc-way-and-transfer-time-group-a-virginia} shows that the proposed method increases the hash verification time, but the total transfer time decreases.
The increase in hash verification time is caused by the overhead of frequent hash function calls and initialization brought by the division of chunks.
The decrease in transfer time was about half the hash verification time of the method calculating the entire hash, indicating that about half of the hash verification time could be executed in parallel during state transfer.

\section{Conclusion}
\label{sec:conclusion}

In this paper, we proposed a state transfer method suitable for geographic SMR.
The proposed method handles the problems caused by the instability of the communication bandwidth specific to geographic SMR, that is, variation and non-uniformity of communication bandwidth.
This is achieved by passively estimating the communication bandwidth between replicas and dynamically adjusting the amount of state that each transfer replica sends to a recovery replica according to their communication bandwidth.
The evaluation results showed that the proposed method can deal with the communication bandwidth variations and reduce the state transfer time by up to 47\% compared to the existing method.
As future work, we believe that the proposed method can be used for the dynamic relocation of replicas according to the communication channel state and the location of clients.
In this way, we can realize efficient SMR that can flexibly respond to various changes in the environment from time to time.

\small

\bibliographystyle{IEEEtran}
\bibliography{collection.bib}

\end{document}